# Galactic astronomy with AO: Nearby star clusters and moving groups

T. J. Davidge[a]

[a]Dominion Astrophysical Observatory, Victoria, BC Canada

## ABSTRACT

Observations of Galactic star clusters and objects in nearby moving groups recorded with Adaptive Optics (AO) systems on Gemini South are discussed. These include observations of open and globular clusters with the GeMS system, and high Strehl $L'$ observations of the moving group member Sirius obtained with NICI. The latter data fail to reveal a brown dwarf companion with a mass $\geq 0.02 M_\odot$ in an $18 \times 18$ arcsec$^2$ area around Sirius A. Potential future directions for AO studies of nearby star clusters and groups with systems on large telescopes are also presented.

## 1. STAR CLUSTERS AND THE GALAXY

Deep surveys of star-forming regions have revealed that stars do not form in isolation, but instead form in clusters or loose groups (e.g. Ref. 25). This is a somewhat surprising result given that most stars in the Galaxy and nearby galaxies are not seen to be in obvious clusters (e.g. Ref. 31). This apparent contradiction can be reconciled if clusters tend to be short-lived. In fact, the pace of cluster destruction has been measured to be roughly an order of magnitude per decade in age (Ref. 17), indicating that the vast majority of clusters have lifespans that are only a modest fraction of the dynamical crossing-time of the Galactic disk. Clusters likely disperse in response to sudden changes in mass driven by supernovae and stellar winds (e.g. Ref. 25), tidal interactions with nearby giant molecular clouds (e.g. Ref. 24), and/or internal relaxation. Given the comparatively rapid pace of cluster disruption, then it is possible (likely?) that the clusters that are seen today may not be representative of the vast majority of clusters that originally formed.

While the time that a typical star spends in a cluster may only be a modest fraction of its total life span, the natal enviroment can have a profound influence on the nature of the stars that form, and ultimately disperse into the so-called 'field'. For example, cluster environment may affect the ability to form low mass objects, which take longer to collapse than higher mass objects and so may have their early evolution curtailed by – say – close proximity to a massive hot star (e.g. Ref. 16). As clusters virialize, low mass stars are preferentially lost, as they develop higher velocities than more massive cluster members, and so the mass function of a cluster then evolves with time. Interactions between cluster members during virialization may also disrupt all but the most tightly bound binary systems. If the mass distribution of stars in binaries follows that of isolated field stars then the contribution that a cluster makes to the low mass end of the field mass function is tied to the properties of the binaries that are disrupted (e.g. their binding energy) and the density of the cluster environment (e.g. Ref. 26).

Studies of relatively young clusters provide insights into the early stages of cluster evolution, and how the stellar content changes with time. Basic cluster parameters, such as mean age and the age dispersion, the cluster mass function, the frequency of binary and higher order stellar systems, and the degree of cluster disruption can be obtained from imaging observations. There are compelling motivations to conduct surveys of cluster stellar content with wide-field AO systems. These systems deliver their best performance as one moves to longer wavelengths, thereby helping to overcome the heavy extinction that is inherent to star-forming regions and is prevalent throughout the Galactic disk. The infrared is also a good wavelength regime for studying low mass stars and pre-main sequence (PMS) objects, as both tend to have relatively low effective temperatures.

---

Further author information: Email: tim.davidge@nrc.ca, Telephone: 250-363-0047

Discrepancies between observations and models of PMS evolution are expected to be smaller in the NIR than at visible wavelengths (Ref. 4).

The improved angular resolution delivered by AO systems when compared with natural seeing conditions is also useful to resolve multiple objects with small projected separations on the sky that might otherwise appear as a single object. Such blends will appear as erroneous points on CMDs, and confuse efforts to characterize cluster properties. In addition to resolving nearby companions, studies of the circumstellar environments of bright cluster members at moderate to high Strehl may yield clues into the evolution of the host star and insights into the mechanisms that drive the early stages of cluster evolution. That the diffraction limit of large ground-based facilties is many times finer than that of the HST in $K$ is a significant advantage in efforts to detect blends and probe circumstellar environments.

In contrast to open clusters, globular clusters are fossils from a time when conditions in the Galaxy were very different from those at the present-day. Globular cluster formation in the nearby Universe is associated with elevated levels of star-forming activity that is likely triggered by tidal interactions and mergers (e.g. Refs. 2 and 36). The old globular clusters that surround the present-day Milky-Way likely formed in proto-galactic structures that were destroyed and had their contents dispersed by the mergers that left behind the gas from which the Milky-Way disk formed (e.g. Ref.s 3 and 22). The longevity of globular clusters when compared with open clusters likely reflects their large masses and orbits that have kept the present-day population in low density environments for most of their lives. Still, the detection of tidal trails from globular clusters (e.g. Ref. 28) indicates that they are dissipating, and have contributed to the field population.

## 2. GEMS AND OPEN CLUSTERS

Open clusters are found throughout the Galactic Disk, and those that are within a few kpc typically subtend an arcmin or more on the sky. The angular extents of such clusters match the field size of GeMS+GSAOI (Refs. 29 and 27), and four open clusters were observed with this system as part of program GS-2012B-SV-409 (PI: Davidge). Each cluster was observed in $J$, $Ks$, and Br$\gamma$, and only the $J$ and $Ks$ observations are considered here. A full discussion of these data can be found in Refs. 14 and 13.

The data discussed here were recorded during 85%ile IQ conditions with the goal of assessing the performance of GeMS during mediocre seeing. This was done in part because MCAO systems, such as NFIRAOS on the TMT (Ref. 21, are an essential part of the instrumentation complement for 30 meter telescopes, as they will allow these facilities to achieve the $D^4$ advantage over moderately large fields. Given the high cost of instruments on very large telescopes, it is anticipated that MCAO systems will be one of only a handful of instruments available at any given time at these facilities, and so it is likely that there will be pressure to use these systems during less-than-ideal observing conditions. Hence, it is important to characterize the performance of MCAO systems during such conditions.

The natural guide stars (NGSs) used for these observations were distributed in a triangular-shaped asterism across each cluster, which is a configuration that is conducive to the PSF uniformity that is important for photometric measurements and astrometric studies. The level of stability obtained here is demonstrated in Fig. 1, which shows the final GeMS+GSAOI $Ks$ images of NGC 3105. Even though the data were recorded during mediocre conditions, there are only trace signatures of image distortion, indicating that residual anisoplanaticism – while present – is modest in size.

The FWHMs of bright, isolated stars in six different sub-regions are also listed in Fig. 1. While the FWHM significantly exceeds the $Ks$ telescope diffraction limit of $\sim 0.06$ arcsec, the delivered angular resolution is still comparable to what would be obtained from a 4 meter telescope during superb conditions. Also, in spite of the mediocre seeing conditions, the angular resolution is relatively stable across each field, with a $\sim 10\%$ difference in the FWHM measurements. The angular resolution is poorer in $J$ ($\sim 0.18$ arcsec FWHM for NGC 3105) although – as with the $Ks$ data – the FWHM in this filter is also stable to within $\sim 10\%$.

The $(K, J-K)$ CMDs of Haffner 16 and NGC 3105 are shown in Fig. 2. The photometric measurements were made with the PSF-fitting program ALLSTAR (Ref. 33), using a spatially-variable PSF to account for FWHM variations across the field. Despite the 10% variation in FWHM, the CMDs produced with a fixed PSF are not greatly different from those shown in Fig. 2.

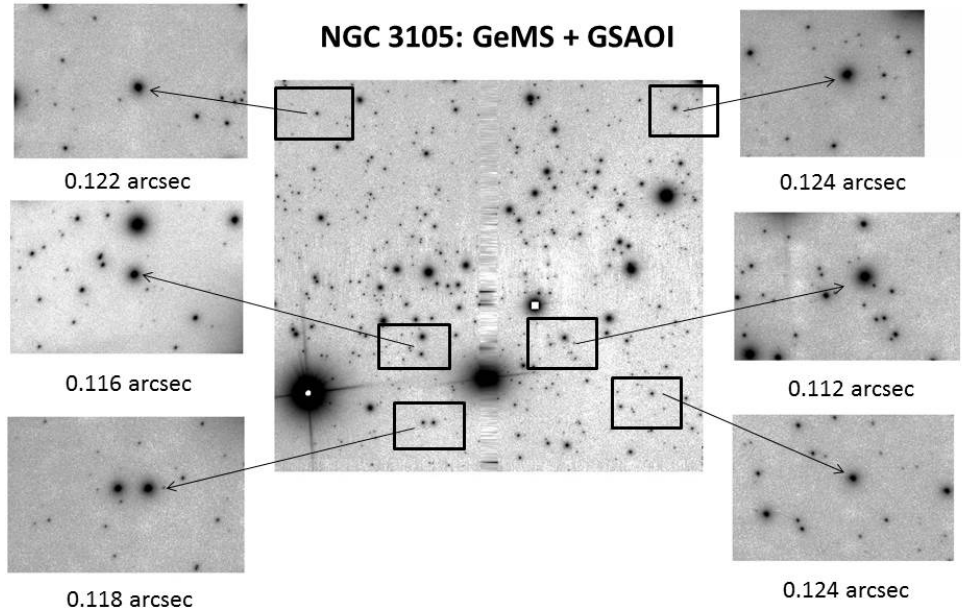

Figure 1. The GeMS+GSAOI $Ks$ image of NGC 3105. The central image covers roughly $85 \times 85$ arcsec$^2$. The streak of bad columns that bisects the field is due to gaps between detectors in GSAOI. The image quality in six sub-fields is also investigated, with the FWHM of the brightest source listed under each sub-panel. Modest evidence of residual anisoplanicity can be seen in the form of elongated images, and the FWHM measurements differ by $\leq 10\%$.

There is significant dispersion along the color axis of each CMD, and this is not a consequence of PSF-variability. Rather, it is due in large part to PMS objects. These contribute the greatest scatter to the Haffner 16 CMD, as this is the younger of the two clusters (10 Myr $vs$ 25 Myr – see below). PMS objects do not follow the main sequence, and – in young clusters like Haffner 16 – can have the same intrinsic brightness as main sequence stars, but very different colors. The presence of circumstellar disks around PMS stars can also contribute to scatter in the PMS sequence. Field star contamination is another source of significant scatter in the NGC 3105 CMD, as star count models predict that at least 40% of objects do not belong to the cluster.

The colored lines in Fig. 2 are isochrones from Ref. 7. These sequences cover evolution from the PMS to past the end of core-helium burning. The distance moduli and reddenings used to place the models on the CMDs have been tuned to match the upper portions of each cluster's main sequence, and so it is by design that the isochrones trace the blue edge near the bright end of the CMDs.

It is heartening that even though the distance moduli and reddenings were largely determined at bright magnitudes, the isochrones amble through the main concentration of points in the CMDs at intermediate and faint magnitudes, where PMS objects dominate. This agreement is satisfying given the uncertainties in models of PMS evolution (e.g. review by Ref. 30). The bowed trajectory of the 4 and 10 Myr isochrones prior to relaxing onto the main sequence demonstrates how PMS stars broaden the color distribution in young clusters. While the majority of stars with intermediate magnitudes in the Haffner 16 CMD fall between the 4 and 10 Myr isochrone, the absence of stars with main sequence like colors that sets in near $K = 15.4$ in Haffner 16 suggests that the cluster has an age near 10 Myr. As for NGC 3105, the 25 Myr isochrone hugs the blue edge of objects on the CMD of that cluster. There is a tight concentration of objects with $J - K \sim 1$ and $K$ between 18 and 19 in the NGC 3105 CMD, which the 25 Myr isochrone matches.

Insights into the dynamical state of these clusters can be gleaned by investigating the spatial distributions of stars that cover different mass ranges. This can be done using the Two Point Angular Correlation Function (TPCF), which is the distribution of angular distances between all pairings of objects in a sample, normalized to that of a random distribution of objects with the same detector geometry. The ratio of the TPCFs of high and low mass stars in Haffner 16 and NGC 3105 are shown in Fig. 3. Corrections have been applied to account for field star contamination using the procedure described by Ref. 14.

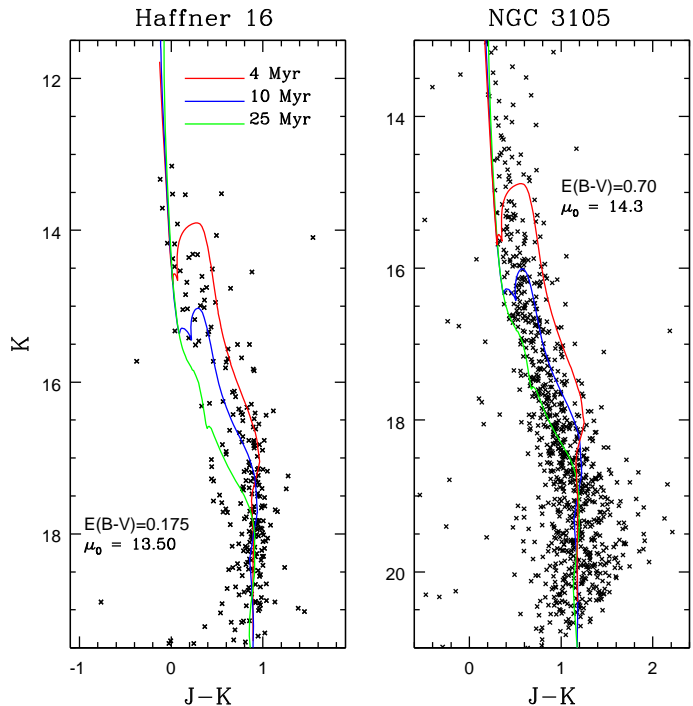

Figure 2. The $(K, J-K)$ CMDs of Haffner 16 and NGC 3105, constructed from GeMS+GSAOI observations. Also shown are solar metallicity isochrones from Ref. 7, that are placed on the CMDs using the reddening and distance modulus quoted in each panel. The dearth of stars with $J-K \sim 0$ when $K > 15.5$ in Haffner 16 is consistent with an age of $\sim 10$ Myr. It is evident that NGC 3105 has an older age, and the 25 Myr sequence passes through the tight sequence with $K$ between 18 and 19 on the NGC 3105 CMD.

If stars in the two mass groups have the same on-sky clustering properties then the ratio of the TPCFs should be constant with separation. The Haffner 16 curve in Fig. 3 is much noisier than the NGC 3105 curve, as it is based on a smaller number of objects. Still, it can be seen that the TPCF ratios for both clusters dip at separations $> 40$ arcsec, indicating that low mass stars in both clusters are less concentrated (i.e. are more diffusely distributed) than high mass stars. This is consistent with the low mass stars being more loosely bound to the cluster than higher mass stars. Both clusters thus appear to be dissipating. It is likely that both clusters will ultimately evolve into diffuse moving groups, possibly like the one that hosts Sirius (Section 4).

## 3. GEMS AND GLOBULAR CLUSTERS

It has long been recognized that globular clusters are important laboratories for characterizing the performance of AO systems (e.g. Ref. 15). Indeed, globular clusters are more richly populated than open clusters, and many are located at high Galactic latitudes, making them less prone to field star contamination. The traditional assumption that clusters have only modest spreads in age and chemical compositions has recently been challenged (e.g. Ref. 23), and there is evidence that dense globular clusters may have experienced star formation over timespans of tens or hundreds of Myr. Still, the old ages of globular clusters mitigates (but does not completely remove) the impact that this might have on their photometric properties when compared with young open clusters.

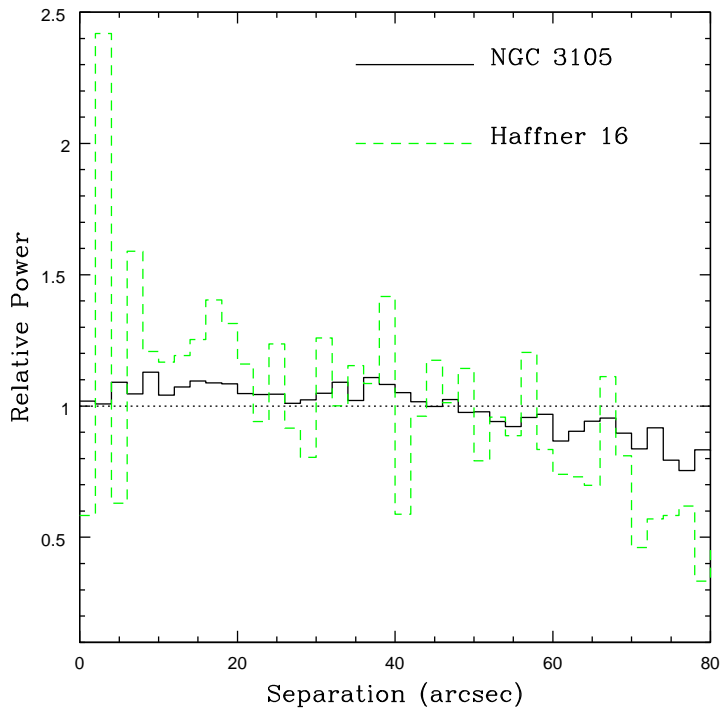

Figure 3. The ratios of the TPCFs of bright and faint stars in Haffner 16 and NGC 3105. The relations have been corrected for field star contamination using the statistical method described by Ref. 14. The bin-to-bin jitter in each curve gives a rough estimate of the uncertainties in the TPCF ratios – there is much more scatter in the Haffner 16 data given the lower number of objects in that cluster. The dotted line marks unity, which is the ratio that would be expected if bright and faint stars in these clusters had similar clustering properties. In both clusters there is a tendency for the TPCF ratio to dip below unity for separations $> 40$ arcsec, indicating that the fainter stars are more diffusely distributed than the brighter stars.

The globular cluster NGC 1851 was observed with GeMS+GSAOI for program GS2012B-SV-406 (PI: McConnachie). The cluster was observed in $J$ and $Ks$, with the primary science goal of measuring the proper motions of cluster members. This information can be used to determine the space motions of the cluster, and when combined with space motions of these clusters, the mass distribution of the Galaxy can then be probed. Faint cluster members that are difficult to distinguish from foreground and background objects using photometric information alone can also be identified with proper motion measurements.

The data were recorded during IQ70%ile conditions. The stability of the PSF is examined in Fig. 4, where the radial behaviour of the FWHM from the center of the cluster is shown. There is a modest radial gradient in FWHM as one moves away from the field center, with the amplitude of the FWHM variation $\geq 0.01$ arcsec. It is worth noting that the dispersion in the FWHM is thus not markedly smaller than what was delivered during IQ85%ile conditions (see Fig. 1).

The stellar content of NGC 1851 can be investigated by combining the GeMS data with existing ACS observations at visible wavelengths, allowing visible/near-infrared CMDs to be constructed. The wide wavelength coverage of such CMDs is particularly useful for constraining stellar properties. The $(K, V - K)$ CMD of NGC 1851 is shown in Fig. 5. There is only a modest dispersion in $V - K$, and the width of the giant branch suggests that the dispersion in $V - K$ is $\sim \pm 0.05$ magnitudes. Effort is being made to reduce this further (McConnachie; private communication).

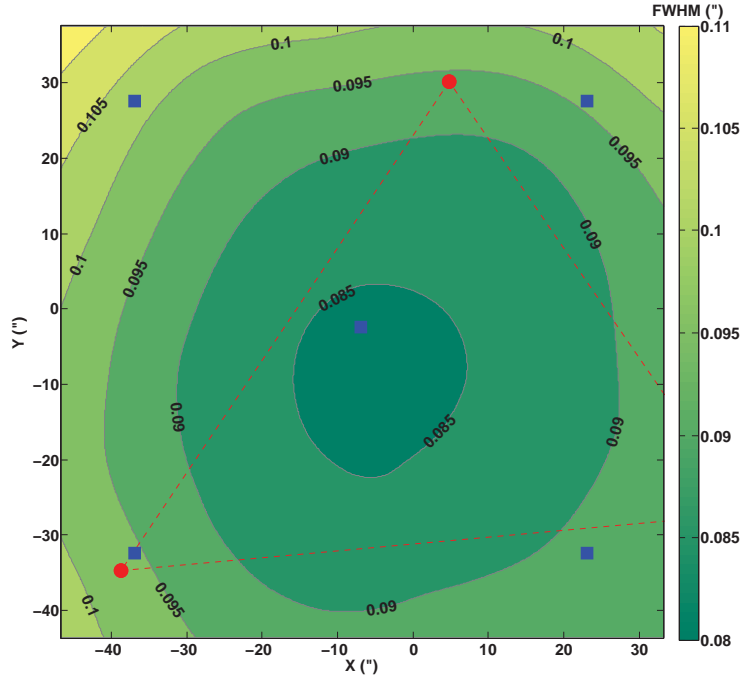

Figure 4. The distribution of FWHM in $K$, measured from GeMS+GSAOI images of NGC 1851. The $\sim 0.015$ arcsec range in FWHM in these data is comparable to that found in the NGC 3105 observations (Fig. 1, even though the NGC 1851 data were recorded during better image quality conditions.

An isochrone from Fiorentino (2014; private communication) that has an old age and a metallicity selected to match that of NGC 1851 is also shown in Fig. 5. There is reasonable agreement between the isochrone and the observations. The GeMS observations track the change in the trajectory of the main sequence near $K \sim 20 - 21$ that is due to a change in the adiabatic gradient near the stellar surface. Additional information about this program and initial science results can be found in Ref. 35.

## 4. THE BINARY FREQUENCY OF BRIGHT STARS IN CLUSTERS AND MOVING GROUPS: A SEARCH FOR THE 'MISSING' COMPANION OF SIRIUS A

Clusters are basic calibrators of stellar properties, and so it is important to understand the nature of the component stars, including the identification of close companions and circumstellar material. Here $L'$ images of one of the brightest members of a nearby moving group – Sirius – are examined. An interesting aspect of the Sirius system, which so far is known to consist of Sirius A and Sirius B, is that there is astrometric evidence for a third, as yet unidentified, member, that may have a mass $\sim 0.05 M_\odot$ (Ref. 5, but see also Ref. 18). Recent surveys have failed to identify such a brown dwarf companion (e.g. Refs. 6 and 34).

The data presented here were obtained with NICI (Ref. 1) on Gemini South as part of program GS-2010A-DD-2 (PI: Davidge). The primary dataset consisted of $L'$ images. The $L'$ filter covers a part of the spectrum that is favorable for the detection of brown dwarfs and warm dust, while at the same time allowing for manageable noise levels from the thermal background. As demonstrated below, it is also possible to obtain high Strehl images with a single NGS system over moderately large angular scales. The observations were recorded with Sirius A as the NGS, and the light from Sirius A was blocked with a 0.65 arcsec radius mask. The total exposure time was 304 seconds, and the data were recorded as 8 observations, each of which consisted of 100 coadds of individual 0.38 sec exposures.

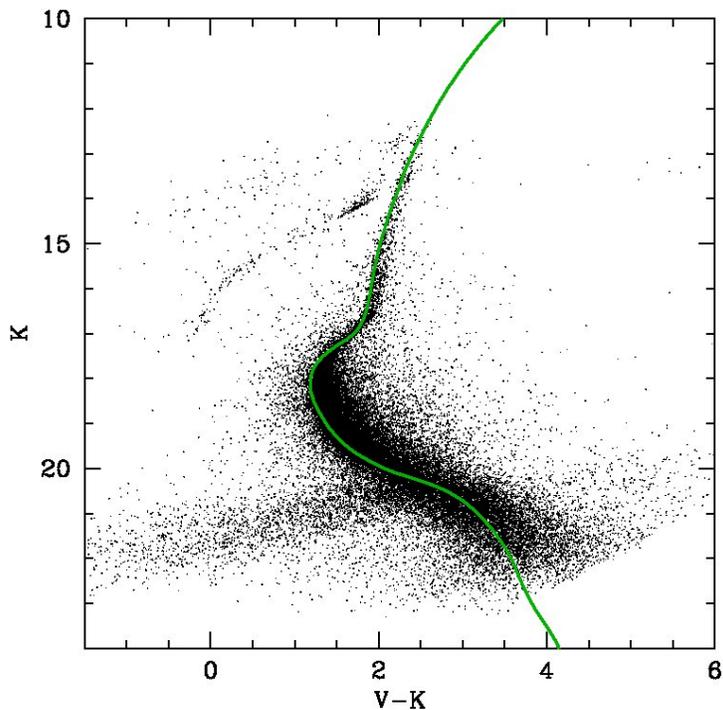

Figure 5. The $(K, V-K)$ CMD of NGC 1851, obtained by combining source catalogues from GeMS and ACS images. The green line is an isochrone of an old population with a metallicity selected to match that of NGC 1851.

The coronographic mask is not completely opaque, and a modest fraction of light from the blocked region is allowed to pass through. The $L'$ image of an $3.5 \times 3.5$ arcsec$^2$ area immediately surrounding Sirius A is shown in the left hand panel of Fig. 6. The 1.3 arcsec diameter region that is occulted by the mask is apparent. Sirius is the central object, and the first diffraction ring is clearly seen. The faint object immediately to the left of Sirius A is a ghost. A halo of light marks the outer boundary of the mask.

The $L'$ image of the full $18 \times 18$ arcsec region imaged by NICI around Sirius A is shown in the right hand panel of Fig. 6. These data have been processed to remove the light from Sirius A by assuming that the PSF is azimuthally symmetric. Residual variations were then removed by subtracting a $9 \times 9$ pixel median boxcar-filtered version of the PSF-subtracted image.

The field imaged by NICI includes the white dwarf Sirius B. Because of the good atmospheric conditions at the time that the data were recorded and the comparatively long wavelength sampled by the $L'$ filter, good correction was delivered for Sirius B, which was $\sim 8$ arcsec distant from Sirius A when the data were recorded. The first diffraction ring around Sirius B is prominent, and there are hints of the second ring at $\sim 0.3$ arcsec radius. The FWHM of Sirius B is 0.09 arcsec, and the azimuthal variation in the light from the diffraction ring due to the secondary mirror support system is also seen. The stable PSF delivered by NICI was previously noted by Ref. 11.

The Sirius B PSF is of interest as it indicates that the entire field is sampled with an angular resolution that is comparable to that of the telescope. It is thus worth noting that – aside from Sirius B – there is only one other point source visible outside of the high-noise area that surrounds Sirius A, and that is a ghost image of Sirius A, which is marked in the image. This ghost is an interesing photometric calibrator, as it is $\sim 3$ magnitudes fainter than Sirius B, and so has an approximate apparent magnitude of $L' = 12$, which corresponds to $M_{L'} \sim 15$ at the distance of Sirius. This $M_{L'}$ is roughly 1 magnitude fainter than measured for a T6.5 dwarf by Ref. 20, and is substantially fainter that expected for a 0.02 $M_\odot$ brown dwarf with an age of 100 Myr (e.g. Ref. 20). Aside from Sirius B and the Sirius A ghost, there are no other stellar sources in the area outside of the high-noise area. These data thus fail to detect the proposed brown dwarf companion to Sirius A. If there is a companion then these data indicate that it was outside of the NICI field at this time.

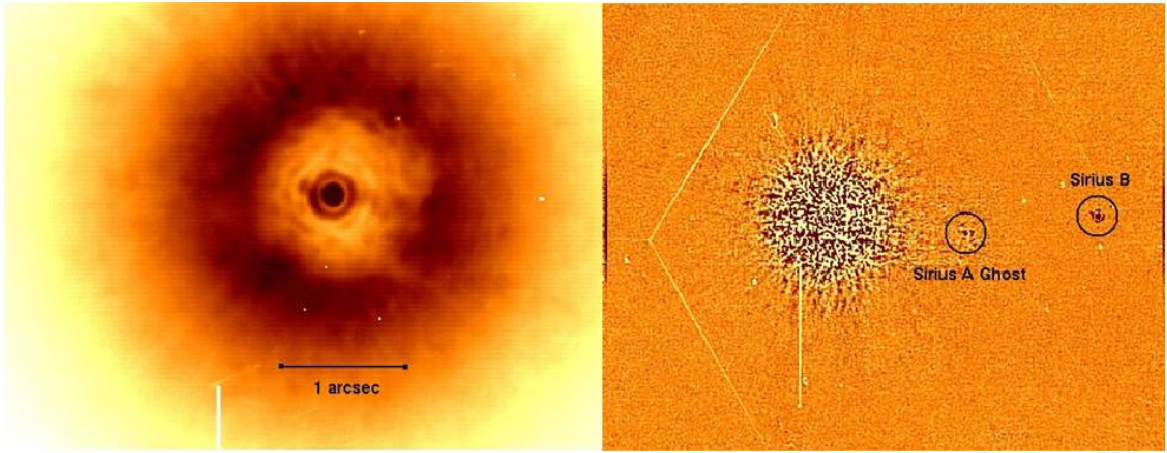

Figure 6. (Left hand panel:) An $L'$ image of a $3.5 \times 3.5$ arcsec$^2$ area surrounding Sirius A, recorded with NICI on Gemini South in 2010. North is towards the top, and east is to the right. Sirius A is at the center of the image, and the 0.65 arcsec radius area blocked by the mask is evident. The object to the immediate left of Sirius A is a ghost. (Right hand panel:) The $18 \times 18$ arcsec field imaged with NICI in $L'$, with artifacts from Sirius A removed using the procedure described in the text. Sirius B and a ghost of Sirius A are both marked. The ghost has a brightness that is at the faint end of the range expected for brown dwarfs, and no other sources of this brightness are seen outside of the high-noise region around Sirius A. Note the presence of the diffraction raing around Sirius B, which demonstrates the large isoplanatic angle when these observations were recorded.

## 5. LOOKING TO THE FUTURE

The next decade will see the commissioning of 30-meter class telescopes equipped with MCAO systems that will offer unique opportunities for studies of Galactic sources. For example, the NFIRAOS system on the TMT will deliver moderate Strehls over a multi-arcmin FOV, yielding higher angular resolutions than are currently available, and permitting objects that are below the detection threshold of current AO systems to be studied.

These facilities will also have extreme AO capabilities. The PFI system that has been proposed for the TMT builds upon the success of GPI on GS. The better angular resolution of a 30 meter telescope will allow PFI to obtain improved contrast at a fixed angular distance than is possible with systems on smaller telescopes. Below, some possible avenues for future work on Galactic clusters are highlighted that could be conducted with systems like NFIRAOS and PFI.

### 5.1 Exploiting the higher angular resolution of NFIRAOS.

The measurement of proper motions from AO corrected images has been pioneered by teams that are exploring the dynamics of objects near the Galactic Center (e.g. Ref. 19). Assuming no systematic aperture-related impediments then the finer angular resolution of the TMT means that proper motions can be measured for objects over a shorter time baseline than is possible with 8 meter facilities. Objects that are more distant or that have smaller proper motions than can be practically studied with an 8 meter telescope will also be within reach of large aperture facilities.

As clusters dissolve they may leave a debris stream in their wake (e.g. Ref. 9), and hints of this has been observed in some open clusters (e.g. Ref. 12). A problem is that the streams are relatively diffuse, and the identification of cluster members is complicated by field star contamination. With an AO system like NFIRAOS it should be possible to obtain proper motions that – when combined with radial velocities – will allow objects that are dissipating from a host cluster to be identified. The velocity dispersions of low mass clusters are a few km sec$^{-1}$, and observations of the Arches cluster by Ref. 10 suggests that astrometry from AO-corrected images can be done to within 1% of the PSF FWHM. Over the time span of a few years then it will be possible to obtain proper motions for objects in and around low mass clusters over a large fraction of the Galactic disk using AO systems on 30 meter telescopes. Low mass members of these clusters should show a larger dispersion in proper motions than the higher mass members if these objects are leaving the cluster.

The finer angular resolution delivered by NFIRAOS will allow relatively close binary systems to be resolved. By measuring brightnesses of the individual components of such systems – as opposed to their blended light – it will be possible to obtain cleaner CMDs. The statistics of close binaries will also help establish the dynamical state of the cluster.

We close this sub-section by noting that the TMT will be a factory churning out newly identified binary systems, that will be discovered serendipitously by the NFIRAOS WFSs during routine target acquisition and set-up. The angular resolutions of the NFIRAOS NGS probes in $J$ is 0.01 arcsec FWHM, and they will guide on sources as faint as $J = 22$. Heretofore unknown binaries (as well as extragalactic objects with modest angular extensions) could be identified by the NFARIOS RTC as objects causing problematic corrections. Binary stars could also be identified visually if the WFS output is displayed to facilitate acquisition.

## 5.2 The use of extreme AO to characterise the circumstellar regions of cluster members.

Stars in clusters are basic calibrators of stellar properties, as they have known distances and ages. In addition to wide-field AO systems, high-Strehl systems are also planned for 30 meter facilities, and these will be useful for probing the environments around cluster members. A limitation is that high Strehl systems like GPI on GS are limited to guide stars with $R > 9$. Similar magnitude restictions will apply to PFI on the TMT if it delivers comparable Strehls.

The circumstellar environments of massive cluster members are potentially important as the winds from these objects may hasten the ejection of interstellar material from a cluster, thereby facilitating its disruption. High Strehl AO systems could be used to probe the circumstellar environment of O stars or M supergiants in young clusters throughout much of the Galactic disk. The spatial extent and density distribution of circumstellar material and signatures of interactions with any surrounding medium could then be used to assess the impact of stellar winds on the surroundings. Candidate objects for such a survey could be selected based on observations by SPITZER and other MIR facilities.

An obvious application of such high Strehl systems is the search for planets and very faint companions. Young clusters are natural places to search for and characterize exo-planets, as planets are brightest during the earliest stages of their evolution, simplifying their detection when compared with older systems.